\documentclass[preprintnumbers,nofootinbib,showkeys,showpacs,amsmath,amssymb]{revtex4}
\usepackage{amsmath,amssymb,graphics,epsfig,subfigure}
\usepackage{color}
\usepackage{multirow}
\usepackage{booktabs}
\usepackage{changes}
\usepackage{footnote}
\usepackage{textcomp}
\usepackage{graphicx}
\usepackage{float}

\usepackage{hyperref}
\hypersetup{colorlinks=true,linkcolor=blue,citecolor=magenta}

\begin{document}
		\renewcommand{\baselinestretch}{1.15}
		
	\title{Are regular black holes from pure gravity classified within the same thermodynamical topology?}
	
	\preprint{}

	\author{Sheng-Wei Wang, Shan-Ping Wu, Shao-Wen Wei \footnote{Corresponding author. E-mail: weishw@lzu.edu.cn}}

	\affiliation{
		$^{1}$Lanzhou Center for Theoretical Physics, Key Laboratory of Theoretical Physics of Gansu Province, and Key Laboratory of Quantum Theory and Applications of MoE, Lanzhou University, Lanzhou, Gansu 730000, China,\\
		$^{2}$Institute of Theoretical Physics, Research Center of Gravitation, and School of Physical Science and Technology, Lanzhou University, Lanzhou 730000, People's Republic of China}
	
	\begin{abstract}
	Regular black holes, which avoid the essential center singularities, can be constructed through various methods, including nonlinear electrodynamics and quantum corrections. Recently, it was shown that via an infinite tower of higher-curvature corrections, one can obtain different regular black hole solutions in any spacetime dimension $D\geq 5$. Utilizing the concept of thermodynamical topology, we examine these black holes as topological thermodynamic defects, classifying them into distinct topological categories based on their generalized free energy. We find that the Hawking temperature of the black hole has at least one zero point at the small horizon radius limit. Under this fact, the regular black holes generated through the purely gravitational theories exhibit universal thermodynamical behaviors, strongly suggesting they belong to the same topological class. We presents a comprehensive analysis of these properties, providing a clearer understanding of the fundamental nature of regular black holes and their classification within the framework of thermodynamical topology.

	\end{abstract}

		\keywords{ Regular black hole, thermodynamical topology}
	\pacs{}
	
	\maketitle
	\newpage
	
		\section{Introduction}\label{intro}
	Black hole thermodynamics is an advanced field of study that investigates the intricate relationship between thermodynamic principles and black hole characteristics. This research interests emerged from the groundbreaking recognition that black holes exhibit entropy and temperature, parameters conventionally linked to thermodynamic systems~\cite{Hawking:1975ParticleCreation,Hawking:1976BHTherm,Bardeen:1976FourLaws,Gibbons:1976ActionIntegral}. Despite significant advancements in exploring black holes, it seems that we still need further means to study them more thoroughly and clearly. To enhance our understanding and characterization of different black hole systems, the concept of thermodynamical topology has been effectively introduced~\cite{Wei:2022dzw}.
	This method views black hole solutions as topological thermodynamic defects, analyzing their topological numbers based on the asymptotic behavior of the generalized free energy, and classifies nearly all black hole systems into three distinct categories: +1, 0, and -1. This groundbreaking work has offered new insights into the fundamental nature of black holes and gravity. The topological approach introduced in Ref.~\cite{Wei:2022dzw} has gained significant popularity due to its wide applicability and efficiency of study. So far, this method has been applied to studies of rotating AdS black holes~\cite{Wu:2023sue,Wu:2022whe} and Lovelock AdS black holes in AdS spacetime~\cite{Bai:2022klw, Liu:2022aqt}. In addition, many other black holes have been studied, such as quantum BTZ black holes, Born-Infeld black holes, dyonic AdS black holes, and many more~\cite{Wu:2024txe,Chen:2023elp,Chen:2024sow,Fang:2022rsb,Chatzifotis:2023ioc,Wei:2023bgp,Du:2023wwg,Chen:2023ddv,Rizwan:2023ivp,Hazarika:2024dex}.
		
	The singularity problem of black holes, as a general prediction of general relativity, signifies the limitations of this theory~\cite{HawkingLargeScale,Senovilla:1998oua, Penrose:1964wq}. In order to address this issue, many fruitful approaches have emerged. Black holes that lack essential singularities but possess only coordinate singularities are termed regular black holes~\cite{Lan:2023cvz}. The study of regular black holes can be traced back to the pioneering work of Sakharov and Gliner, who proposed that essential singularities could be circumvented by substituting the vacuum with a vacuum-like medium characterized by a de Sitter metric~\cite{Sakharov:1966aja,Gliner:1966}.

The early model of regular black hole was developed by Bardeen by making the mass of the Schwarzschild black hole as a position-dependent function~\cite{Bardeen:1968}, which is now referred to as the Bardeen black hole. Ayon-Beato and Garcia provided the first explanation of the Bardeen black hole \cite{Ayon-Beato:2000mjt} that they are described by the Lagrangian with nonlinear electrodynamics source. The solution to the Einstein equations, coupled with the energy-momentum tensor associated with the magnetic field strength, corresponds to a self-gravitating magnetic monopole charge~\cite{Ayon-Beato:2000mjt}. Later, many other regular black hole solutions were obtained by making use of the nonlinear electrodynamics~\cite{Ayon-Beato:1998hmi,Bronnikov:2000vy,Bronnikov:2000yz,Ayon-Beato:2004ywd,Dymnikova:2004zc,Berej:2006cc,Balart:2014jia,Fan:2016rih,Bronnikov:2017sgg,Junior:2023ixh}.

    Recently, Bueno, Cano, and Hennigar introduced a novel method that presents the first instances of regular black hole solutions within purely gravitational theories, requiring no specific fine-tuning or constraints among the relevant parameters~\cite{Bueno:2024dgm}. Their work unveiled innovative families of regular black holes in spacetime dimensions of larger than five. These black holes introduce modifications to the conventional Schwarzschild black hole and belong to the class of quasi-topological gravities, a well-established category of metric theories of gravity with higher-curvature terms. Notably, these modifications are advantageous for investigating black hole solutions as their equations of motion simplify to second order when applied to static and spherically symmetric metrics~\cite{Bueno:2024dgm,Oliva:2010eb,Myers:2010ru,Dehghani:2011vu,Ahmed:2017jod,Cisterna:2017umf}. Thus far, Ref.~\cite{Bueno:2024dgm} has inspired a series of remarkable studies utilizing its proposed approach~\cite{Konoplya:2024hfg,DiFilippo:2024mwm,Konoplya:2024kih,Ma:2024olw,Bueno:2024qhh,Zhang:2024ljd,Estrada:2024moz,Bueno:2024eig,Bueno:2024zsx}.

	We carefully study the thermodynamics of regular black holes constructed using the method of Bueno, Cano, and Hennigar~\cite{Bueno:2024dgm} and found that they exhibit certain universal properties. We conjecture that regular black holes constructed by this method share some common thermodynamic features, classifying them as the same type of black hole. Is there a simple and intuitive way to illustrate this? Fortunately, through the analysis of the general case, we conclude that these black holes can reach zero temperature. By focusing on cases where the Hawking temperature has only one single zero point, our analysis in asymptotic limits shows that thermodynamic topology not only effectively describes their thermodynamic properties but also confirms, from a topological perspective, that they belong to the same category. Moreover, this category is actually $W^{0+}$ topological classification given in Ref. \cite{Wei24}, through which one can find that the the innermost and outermost black hole states are local thermodynamical stable and unstable. In particular, at low temperature limit, there could be the unstable large and stable small black holes appeared in pair. These result provide a clearer and more comprehensive understanding of the thermodynamic properties of these regular black holes, serving as the motivation for this paper.
	
	The paper is organized as follows. In Sec.~\ref{Sec_RegularBlackHolefromPG}, we review the construction of regular black holes using pure gravity methods. In Sec.~\ref{Sec_GeneralDisscusion}, we analyze their thermodynamical topology and draw our main universal conclusions that these regular black holes belong to one topological class. Two examples to validate these results, including specific details about the thermodynamical topology are given in Sec.~\ref{Sec_twoexamples}. Finally, we summarize and discuss our results. The appendix provide some mathematical details.

\section{Regular Black Holes from Pure Gravity}\label{Sec_RegularBlackHolefromPG}

Quasi-topological gravity is especially effective for analyzing black hole solutions, and they exist across all curvature orders in dimensions $D\geq 5$~\cite{Oliva:2010eb,Myers:2010ru,Dehghani:2011vu,Ahmed:2017jod,Cisterna:2017umf}. This family of theories is extensive enough to serve as a foundational set of polynomial densities for expanding general relativity within the framework of effective field theory. As a result, they offer a robust framework for exploring the influence of higher-curvature corrections near black hole singularities. In this section, we would like to give a brief review for such quasi-topological gravity.
		
In quasi-topological gravity theory, the action is
	\begin{equation}
	I_{\mathrm{QT}}=\frac{1}{16\pi G}\int\mathrm d^Dx\sqrt{|g|}\left[R+\sum_{n=2}^{n_{\mathrm{max}}}\alpha_{n}\mathcal{Z}_{n}\right],
	\label{eq_action_QT}
	\end{equation}
where $\alpha_n$ are arbitrary coupling constants with dimensions of $\mathrm{length}^{2(n-1)}$, and $\mathcal{Z}_{n}$ are the quasi-topological densities~\cite{Oliva:2010eb,Myers:2010ru,Dehghani:2011vu,Ahmed:2017jod,Cisterna:2017umf}. Here $n_{\mathrm{max}}$ is an integer greater than or equal to 2, and under certain conditions, it can approach positive infinity. Then, let us consider the line element for the $D$-dimensional ($D\geq5$) static and spherically symmetric black hole
\begin{equation}
		ds^2=-N(r)^2f(r)dt^2+\frac{dr^2}{f(r)}+r^2d\Omega_{D-2}^2.
\end{equation}
To simplify the procedure, one can introduce~\cite{Bueno:2024dgm}
	\begin{equation}
		\psi(r)\equiv \frac{1-f(r)}{r^2},
		\label{eq_phi}
\end{equation}
and series function
	\begin{equation}
		h(\psi)\equiv \psi+\sum_{n=2}^{n_{\max}}\alpha_{n}\psi^{n}.
		\label{eq_h_psi_series}
\end{equation}
Then, the equations of motion derived from the action~\eqref{eq_action_QT} can be expressed as
\begin{equation}
		\frac{dN}{dr}=0, \quad\frac d{dr}\left(r^{D-1}h(\psi(r))\right)=0,
\end{equation}
and its solution reads~\cite{Bueno:2024dgm}
	\begin{equation}
	N(r)=1,\quad	h\left(\psi(r)\right)=h\left(\frac{1-f(r)}{r^2}\right)=\frac \mu{r^{D-1}},
		\label{eq_h_psi}
	\end{equation}
	where $\mu$ is an integration constant which is proportional to the ADM mass of the solution. Note that if
$\alpha_2, \alpha_3,...,\alpha_n$ are given, then the metric function $f(r)$ can be obtained by solving Eq.~\eqref{eq_h_psi}.

\begin{figure}
	\begin{center}
		{\includegraphics[width=10cm]{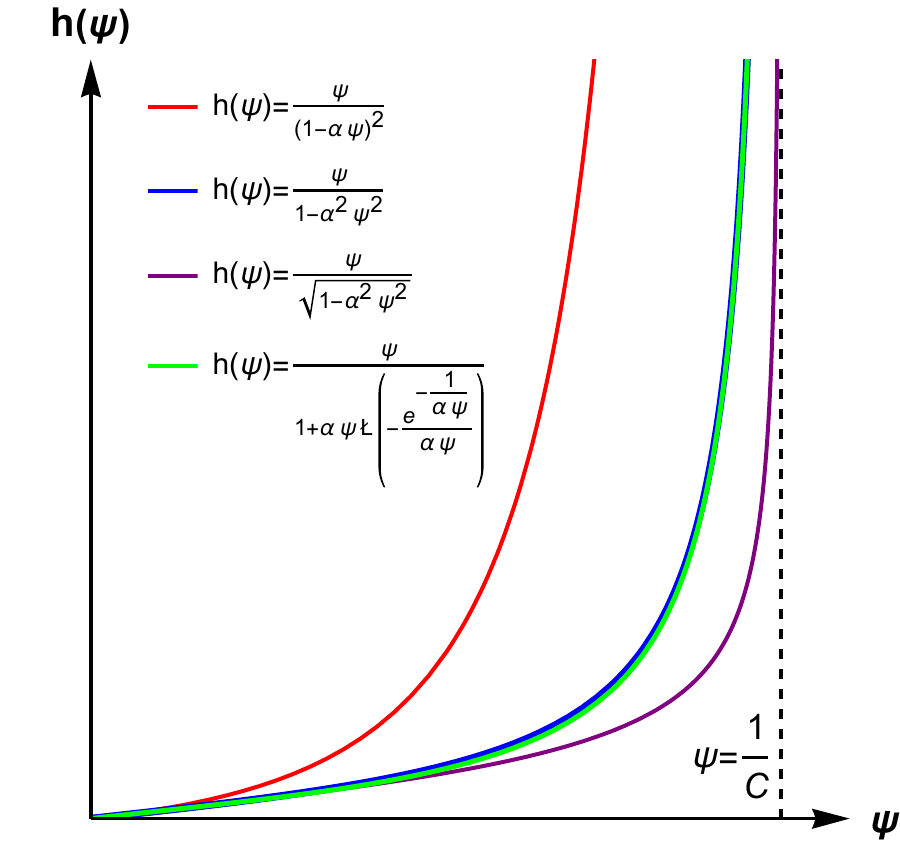}}
	\end{center}
\caption{$h(\psi)-\psi$ diagram of the four regular black holes given in Ref.~\cite{Bueno:2024dgm,Konoplya:2024kih}. The definition of the function $\L$ can be found in Eq.~\eqref{Lambert}. The lines in four different colors represent the plots of $h(\psi)$ corresponding to four different $h(\psi)$ configurations. Here, $\alpha=1$ is chosen. It is easy to prove that the radius of convergence is the same in all four cases.}\label{FIG1}
\end{figure}

Expanding this solution around $r=0$, one has
\begin{equation}
f(r)=1-\left(\frac{m}{\alpha_{n_{\mathrm{max}}}}\right)^{1/n_{\mathrm{max}}}r^{2-(D-1)/n_{\mathrm{max}}}+\cdots.
\end{equation}
 If $n_{\mathrm{max}}$ is finite, a curvature singularity occurs at $r=0$. However, as $n_{\mathrm{max}}\rightarrow\infty$, the metric near $r=0$ approaches that of a de Sitter universe with a cosmological length scale determined by $m/\alpha_{n_{\mathrm{max}}}$, which resolves the singularity. To ensure that the black holes with regular interiors exist as solutions to the full theory, the following conditions are sufficient,
	    \begin{equation}
	\alpha_n\geq0\mathrm{~}\forall\; n,\quad\lim_{n\to\infty}(\alpha_n)^{\frac1n}=C>0.
	\label{C}
	\end{equation}
	Here, the condition that $\alpha_n$ is positive ensures the monotonicity of $h(\psi)$ for $\psi > 0$, while the limit $\lim_{n\to\infty}(\alpha_n)^{\frac{1}{n}} = C > 0$ determines the radius of convergence of the series~\eqref{eq_h_psi_series} as $\psi_0 = 1/C$. At $\psi = \psi_0$, the function $h(\psi)$ diverges. These conditions imply that $h$ is a smooth, bijective mapping from $(0, 1/C)$ to $(0, \infty)$, ensuring the existence of an inverse function for $h(\psi)$. To clearly illustrate the divergence of $h(\psi)$, we show its behavior under various conditions in Fig.~\ref{FIG1}. Our calculations reveal that the divergence radius for the four cases is $\psi=1/C=1/\alpha$. These properties are crucial for the following analysis and will significantly influence the thermodynamic characteristics of the black holes under consideration, which will be discussed in detail subsequently.

\section{Thermodynamic topology}\label{Sec_GeneralDisscusion}

Topological methods have emerged as a powerful tool for studying black hole thermodynamics, offering direct insights into their fundamental properties and enabling classification based on the topological invariants~\cite{Wei:2022dzw}. This approach provides a novel perspective on black hole physics, allowing us to understand their behavior from a topological viewpoint.  Quasitopological gravity, an important class of gravitational theories discussed earlier, plays a key role in constructing various regular black hole solutions. Our goal is to further classify these solutions by using the framework of thermodynamic topology.

For black hole, the geometry near its event horizon is of critical importance. To analyze the horizon condition, one needs to examine $f(r)=0$. From Eq.~\eqref{eq_phi}, we find that
	\begin{equation}
		\psi_+\equiv \psi(r_+)=\frac{1}{r_+^2}.
		\label{eq_psi=1/r_+^2}
	\end{equation}
Here, $r_+$ denotes the largest root of the equation $f(r)=0$, which corresponds to the radius of the event horizon. The term $\psi_+$ indicates the value of $\psi(r)$ at $r=r_+$. Simultaneously, Eq.~\eqref{eq_h_psi} gives
	\begin{equation}
		h(\psi_{+})=\frac{\mu}{r_{+}^{d-1}}.
	\end{equation}
	
Since $h$ is closely related to the black hole metric, analyzing the behavior of $h$ is crucial for further studying the thermodynamic properties of black holes. Firstly, as discussed earlier, $h$ is a smooth, monotonically increasing function within its domain $(0,1/C)$. Secondly, when $\psi_+\rightarrow0$, we have $r_+\rightarrow\infty$ and $h(\psi_+)\rightarrow0$; when $\psi_+\rightarrow1/C$, $r_+\rightarrow\sqrt{C}$  and $h(\psi_+)\rightarrow\infty$, as listed in Table~\ref{Table1}.
\begin{table}[H]
	\centering
	\resizebox{7cm}{!}{
		\begin{tabular}{cccc}
			\hline
			\ \ \ \ \ \ $\psi_+$ \ \ \ \ \ \ & \ \ \ \ \ \ $r_+$ \ \ \ \ \ \ &  \ \ \ \ \ \ $h(\psi_+)$  \ \ \ \ \ \  \\
			\hline
			$ \psi_+\rightarrow 0 $  & $r_+\rightarrow\infty $ & $h(\psi_+)\rightarrow0$  \\
			$ \psi_+\rightarrow 1/C $  & $r_+\rightarrow\sqrt{C} $ & $h(\psi_+)\rightarrow\infty $ \\
			
			\hline
	\end{tabular}}
	\caption{Boundary behaviors of $h(\psi_+)$.}\label{Table1}
\end{table}

Thus, $r_+$ should belong to the range $(\sqrt{C},\infty)$. The value of the lower bound $\sqrt{C}$ arises from the analytic domain of function $h$ itself, as shown in Fig.~\ref{FIG1}. Another point to note is that $\sqrt{C}$ obtained here is not necessarily the lower bound of $r_+$, which is because that Eq.~\eqref{eq_psi=1/r_+^2} only indicates that $f(r_+)=0$, while not implies $r_+$ is the largest root.

Drawing upon the conclusions from Ref.~\cite{Bueno:2024dgm}, one derives the thermodynamic quantities of the regular black holes as follows:
\begin{equation}
			M=\frac{(D-2)\Omega_{D-2}r_{+}^{D-1}}{16\pi G}h(\psi_{+}),
			\label{eq_M}
\end{equation}
\begin{equation}
			\begin{aligned}T=\frac{1}{4\pi r_+}\left[\frac{(D-1)r_+^2h(\psi_+)}{h'(\psi_+)}-2\right]\end{aligned},
			\label{eq_T}
\end{equation}
\begin{equation}
			S=-\frac{(D-2)\Omega_{D-2}}{8G}\int\frac{h'(\psi_{+})}{\psi_{+}^{D/2}}\mathrm{d}\psi_{+},
			\label{eq_S}
\end{equation}
where $M$, $T$, $S$ are the ADM mass, Hawking temperature, and Wald entropy~\cite{Wald:1993nt}, respectively. The volume of the $D$-dimensional unit sphere reads
\begin{equation}
			\Omega_{D}=\frac{2 \pi^{\frac{D+1}{2}}}{\Gamma(\frac{D+1}{2})}.
\end{equation}
Here, we continue to analyze the general expression of the Hawking temperature. Substituting Eqs.~\eqref{eq_h_psi_series} and \eqref{eq_psi=1/r_+^2} into Eq.~\eqref{eq_T}, one will have
\begin{equation}
		T=\frac{1}{4\pi r_+}\left[\frac{(D-1)\left(1+\sum_{n=2}^{n_{\max}}\alpha_{n}\psi_+^{n-1}\right)}{1+\sum_{n=2}^{n_{\max}}n\alpha_{n}\psi_+^{n-1}}-2\right].
		\label{eq_T_subsitutingh(psi)2}
\end{equation}
Here, we note that as $r_+\to\infty$ and $\psi_+\to 0$, the two series $\sum_{n=2}^{n_{\max}}\alpha_{n}\psi_+^{n-1}$ and $\sum_{n=2}^{n_{\max}}n\alpha_{n}\psi_+^{n-1}$ in Eq.~\eqref{eq_T_subsitutingh(psi)2} tend to zero. Consequently, we obtain
\begin{equation}
	\lim_{r_+ \to \infty} T(r_+)\to\frac{1}{4\pi r_+}(D-3).
\end{equation}
Clearly, since $D\geq5$, the temperature $T$ will approach $0^+$ for large horizon radius. Conversely, as $r_+\rightarrow\sqrt{C}$, $\psi_+$ reaches its radius of convergence. Based on the analysis results in Appendix~\ref{appendixprove}, we conclude that when $r_+$ is sufficiently close to $\sqrt{C}$, the temperature $T$ can become negative. Thus, on the interval $(\sqrt{C},\infty)$, $T(r_+)$ will have at least one zero and must have an odd number of zeros if the number of zero points is finite. Considering the complexity of multiple horizon scenarios, we mainly focus on the simplest case where $T(r_+)$ has a single zero in the following discussion. It is worth noting that when $T(r_+)<0$, $r_+$ will represent the inner horizon rather than the outer horizon, losing its original interpretation. To prevent this, $r_+$ gains a new lower bound $r_{+\mathrm{min}}$, given by $T(r_+)=0$. This new lower bound is greater than the previously mentioned lower bound $\sqrt{C}$, establishing $r_{+\mathrm{min}}$ as the true lower bound of $r_+$.

\subsection{Topological approach}

To describe the thermodynamical topology of black holes, we consider the generalized free energy of a regular black hole given by
\begin{equation}
			F=M-\frac{S}{\tau},
			\label{eq_F}
\end{equation}
where $\tau$ represents an extra variable, which can be regarded as the inverse temperature of the cavity surrounding the black hole. When $\tau=1/T$, the free energy becomes on-shell. We then examine the vector $\phi$, defined as
\begin{equation}
			\phi=\begin{pmatrix}\frac{\partial F}{\partial r_+},-\cot\Theta\csc\Theta\end{pmatrix},
			\label{phi}
\end{equation}
where $r_+\in \left(0,\infty\right)$ and $\Theta\in\left(0,\pi\right)$. The component $\phi^\Theta$ diverges at $\Theta=0$ and $\Theta=\pi$, implying that the vector $\phi$ points outward.
		
Following Duan's theory of $\phi$-mapping topological currents~\cite{YSD,YSD1}, the topological current can be constructed as
\begin{equation}
			j^\mu=\frac{1}{2\pi}\varepsilon^{\mu\nu\rho}\varepsilon_{ab}\partial_\nu n^a\partial_\rho n^b,\quad\mu,\nu,\rho=0,1,2,
\end{equation}
where $\partial_{\nu}=\partial/\partial x^{\nu}$ and $x^{\nu}=(\tau,r_+,\Theta)$. To normalize $\phi$, we use $n^{r}=\frac{\phi^{r}}{\|\phi\|}$ and $n^{\Theta}=\frac{\phi^{\theta}}{\|\phi\|}$. The topological current $j^\mu$ is conserved, satisfying
\begin{equation}
			\partial_\mu j^\mu=0.
\end{equation}
Using the Jacobi tensor $\epsilon^{ab}J^{\mu}(\phi/x)=\epsilon^{\mu\nu\rho}\partial_\nu\phi^a\partial_\rho\phi^b$ and the two-dimensional Laplacian Green's function $\Delta_{\phi^{a}}\ln\|\phi\|=2\pi\delta^{2}(\phi)$, the current $j^\mu$ can be expressed as
\begin{equation}
			j^\mu=\delta^2(\phi)J^\mu\left(\frac\phi x\right).
\end{equation}
Since $j^\mu$ is nonzero only at $\phi^{a}(x^{i})=0$, one can denote its $i$-th solution as $\vec{x}=\vec{z}_{i}$. The density of the topological current is given by
\begin{equation}
			j^0=\sum_{i=1}^N\beta_i\eta_i\delta^2(\vec{x}-\vec{z}_i).
\end{equation}
The positive Hopf index, denoted as $\beta_i$, quantifies the number of loops that $\phi^\alpha$ traces in the vector $\phi$ space as $x^\mu$ circles the zero point $z_i$. Meanwhile, the Brouwer degree, denoted as $\eta_i$, is given by the expression $\eta_i = \mathrm{sign}(J^0(\phi/x)_{z_i})=\pm1$. Within a certain parameter region $\Sigma$, the associated topological number can be determined by
\begin{equation}
			W=\int_{\Sigma}j^0d^2x=\sum_{i=1}^N\beta_i\eta_i=\sum_{i=1}^Nw_i,
\end{equation}
where $w_i$ represents the winding number associated with the $i$-th zero point of $\phi$ contained within $\Sigma$. If loops $\partial\Sigma$ and $\partial\Sigma^{\prime}$ enclose the same zero point of $\phi$, we will have the same winding number. If there are no zero points in the enclosed region, $W=0$. When $\Sigma$ covers part or all of the parameter space, it reveals the local or global topological number, respectively.

\subsection{Thermodynamic number}

To study the thermodynamical topology, we will further analyze the asymptotic behavior of the thermodynamic quantities. To obtain Eq.~\eqref{phi}, we need to differentiate Eq.~\eqref{eq_F} with respect to $r_+$, resulting in \cite{Wei24}
\begin{equation}
		\frac{\partial F}{\partial r_+}=\frac{\partial M}{\partial r_+}-\frac{1}{\tau}\frac{\partial S}{\partial r_+}=\frac{\partial S}{\partial r_+}\left(\frac{\partial M}{\partial S}-\frac{1}{\tau}\right)=\frac{\partial S}{\partial r_+}\left(T-\frac{1}{\tau}\right)
		\label{eq_partialF/partialr_+}.
\end{equation}
Here, we have used the first law of black hole thermodynamics. Further considering the most general expressions~\eqref{eq_M},~\eqref{eq_T},  and~\eqref{eq_S}, we shall obtain the general conclusions.
	
By substituting Eq.~\eqref{eq_psi=1/r_+^2} into Eq.~\eqref{eq_S}, we obtain
\begin{equation}
		S=\frac{(D-2)\Omega_{D-2}}{8G}\int\frac{2h'(\psi_{+})r_+^{-3}}{\psi_{+}^{D/2}}\mathrm{d}r_{+}.
		\label{eq_S_dr_+}
\end{equation}
Here, the prime symbol denotes the first derivative of function $h$ with respect to $\psi$, which is given by
\begin{equation}
		h'(\psi)=1+\sum_{n=2}^{n_{\max}}n\alpha_{n}\psi^{n-1}.
		\label{eq_h_psi_series_prime}
\end{equation}
Thus, we can derive
\begin{equation}
		\frac{\partial S}{\partial r_+}=\frac{(D-2)\Omega_{D-2}}{8G}\frac{2h'(\psi_{+})r_+^{-3}}{\psi_{+}^{D/2}}.
\end{equation}
Clearly, it follows that when $r_+>0$, it is always true that
\begin{equation}
		\frac{\partial S}{\partial r_+}>0.
\end{equation}
	
\begin{figure}[H]
		\begin{center}
			{\includegraphics[width=9cm]{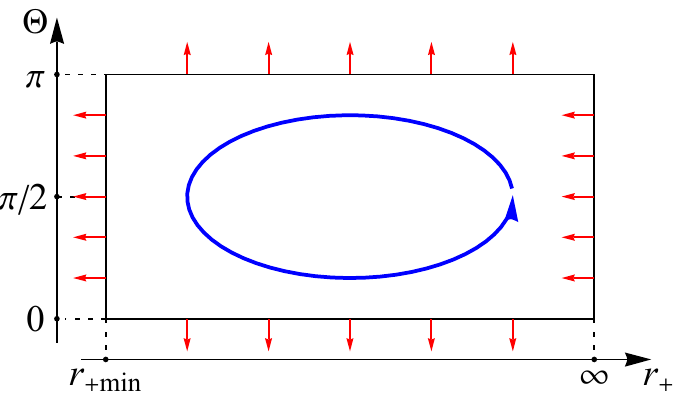}}
		\end{center}
		\caption{$\Theta-r_+$ schematic diagram of the regular black holes. According to our discussion of the general case, the $\Phi-r_+$ diagram of such black holes exhibits certain common features. The vector direction on the boundaries is fixed, thus they have the same winding number. This preserves the technical terminology in a way that should be suitable for your context in black hole physics. }\label{FIG2}
	\end{figure}

Note that in Eq.~\eqref{eq_partialF/partialr_+}, $\tau$ is positive. Considering the conclusions we reached earlier, we can derive that when $r_+\rightarrow r_{+min}$, $T\rightarrow0$, and $T-1/\tau<0$; when $r_+\rightarrow\infty$, $T\rightarrow0$, and $T-1/\tau<0$. Therefore, by combining the signs of two terms $\frac{\partial S}{\partial r_+}$ and $\left(T-\frac{1}{\tau}\right)$ in Eq~\eqref{eq_partialF/partialr_+}, we can conclude that when $r_+$ takes $r_{+min}$ or tends to infinity, $\frac{\partial F}{\partial r_+}$ is less than zero, as listed in Table~\ref{Table2}.
\begin{table}[H]
	\centering
	\resizebox{9cm}{!}{
		\begin{tabular}{cccc}
			\hline
			\ \ \ \ \ \ $r_+$ \ \ \ \ \ \ & \ \ \ \ \ \ $T$ \ \ \ \ \ \ &  \ \ \ \ \ \ $T-1/\tau$ \ \ \ \ \ \ & \ \ \ \ \ \ $\frac{\partial F}{\partial r_+}$ \ \ \ \ \ \  \\
			\hline
			$ r_+\rightarrow r_{+min} $  & $T\rightarrow0^- $ & $T-1/\tau<0$ & $\frac{\partial F}{\partial r_+}<0$ \\
			$ r_+\rightarrow\infty $  & $T\rightarrow0^+ $ & $T-1/\tau<0$ & $\frac{\partial F}{\partial r_+}<0$ \\
			
			\hline
	\end{tabular}}
	\caption{Boundary behavior of $\frac{\partial F}{\partial r_+}$.}\label{Table2}
\end{table}
This implies that in the $\Theta-r_+$ diagram, as $r_+$ approaches the lower bound and infinity, the direction of the vector $\phi$ remains consistent (leftwards). At the same time, we will note that when $\Theta$ approaches zero, the direction of the arrows in the $\Theta-r_+$ diagram remains fixed (downwards); and when $\Theta$ approaches $\pi$, the direction of the arrows in the $\Theta-r_+$ diagram also remains fixed (upwards). The behaviors of the above vectors at the boundaries have been illustrated in  Fig.~\ref{FIG2}. Employing with above results, we can derive some universal properties on a global scale. If we consider a sufficiently large loop that encompasses all the zeros of $\frac{\partial F}{\partial r_+}$ in the $\Theta-r_+$ diagram, the global topological number can be obtained as zero
\begin{equation}
	W=0.
\end{equation}
	
	Since all the discussions derived thus far are general, the black holes obtained under the aforementioned considerations can be classified as the same type with a topological number $W=0$. Combining with the vector direction on the boundaries, we find their topological class actually is $W^{0+}$ suggested in Ref. \cite{Wei24}. Meanwhile, we can also derive the result that the small and large black hole states are stable and unstable from the general property of $W^{0+}$. This results imply that the black hole solutions constructed from the pure gravity approach outlined above belong to the same topological class, which indicates that certain inherent topological properties are embedded in the method of constructing such regular black holes. Specifically, the geometric and topological constraints of these construction techniques ensure that universal features are preserved across various gravitational backgrounds.

	A key requirement in our analysis is the condition that the temperature $T(r_+)$ has a single zero point. This constraint directly influences the topological classification and physical interpretation of the black hole states. In black hole thermodynamics, topological numbers $W=1$ and $W=-1$ correspond to stable and unstable black hole solutions, respectively~\cite{Wei:2022dzw}. A global topological number $W=0$ thus indicates a balance between stable and unstable black hole states. This balance arises because the positive topological numbers of stable states are exactly canceled by the negative topological numbers of unstable states, leading to a net topological number of zero.
	
	This equilibrium is not coincidental but is deeply rooted in the inherent symmetry and geometric characteristics of the gravitational field. It reveals an underlying topological structure, wherein perturbations in the gravitational field within a multidimensional spacetime background generate equal numbers of stable and unstable solutions, thereby maintaining a global topological balance. This phenomenon provides a novel perspective on the construction and evolution of regular black holes, highlighting their significance within the framework of generalized gravity theories.
	
	In the subsequent discussion, we will provide specific examples to verify our perspective, exploring the topological structure and thermodynamic properties of these regular black hole solutions.

\section{Two regular black hole solutions}\label{Sec_twoexamples}

In the previous section, we consider the thermodynamical topology for the general case and find that the total topological number vanishes. In this section, we will take two specific black hole solutions as specific examples to verify our above results.

\subsection{Regular black hole with $\alpha_n=n\alpha^{n-1}$}\label{Subsec_example1}

\begin{figure}[H]
	\begin{center}
		\subfigure[\label{FIG3a}]{\includegraphics[width=5cm]{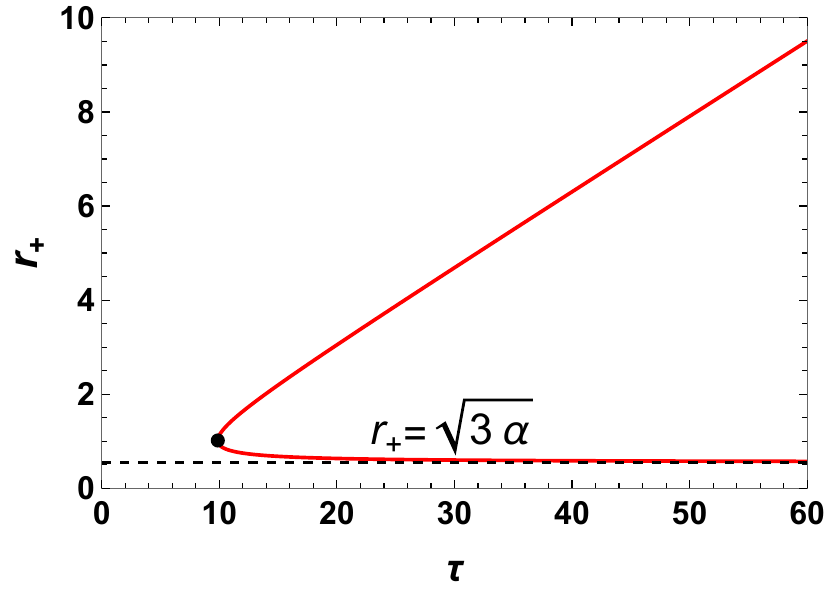}}
		\subfigure[\label{FIG3b}]{\includegraphics[width=5cm]{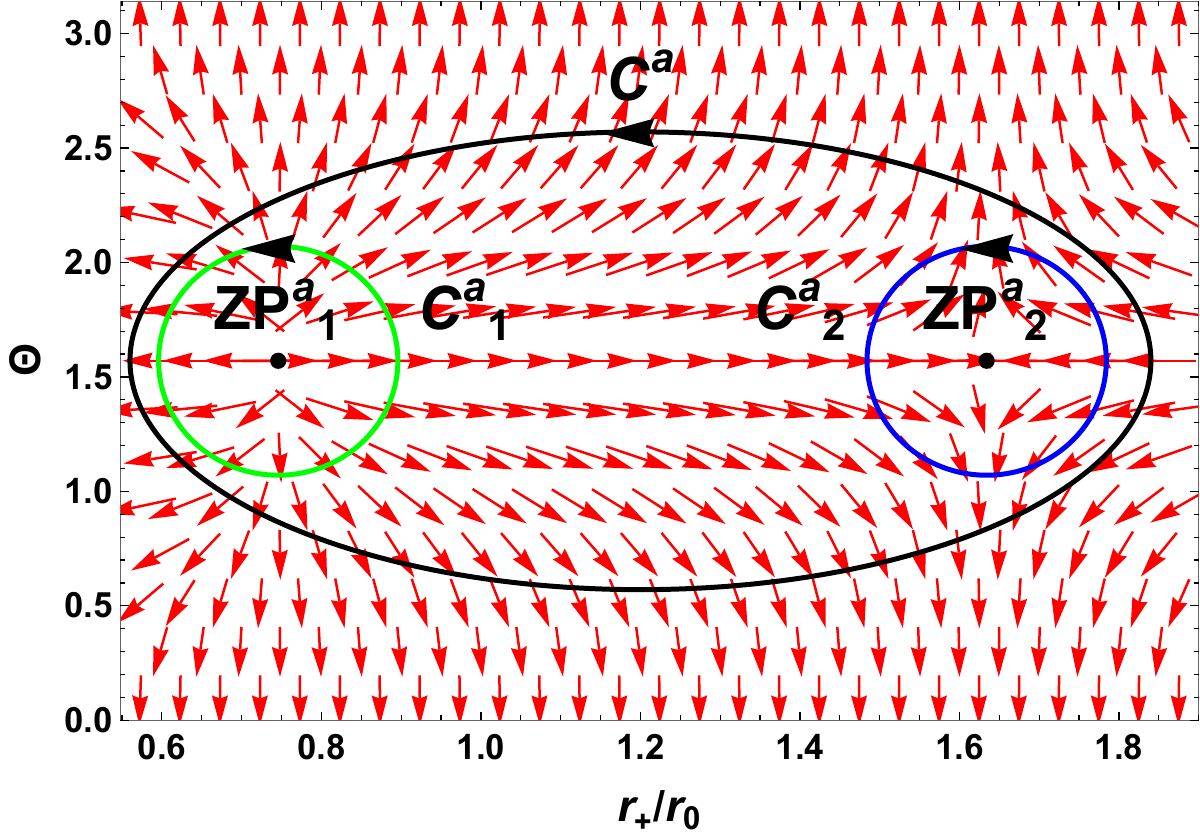}}
		\subfigure[\label{FIG3c}]{\includegraphics[width=5cm]{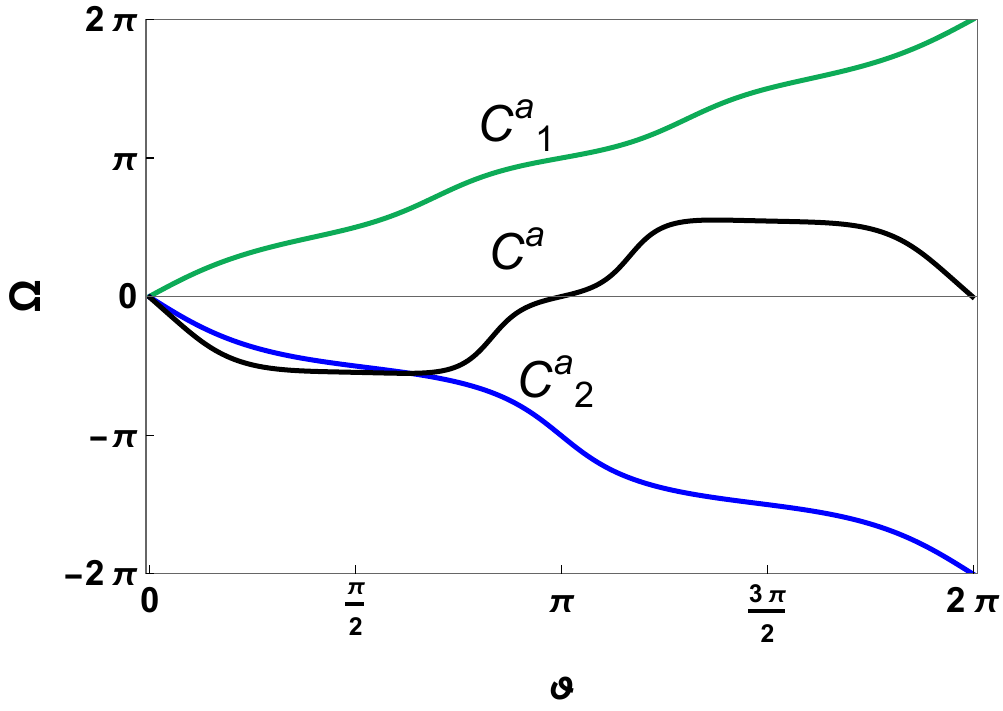}}
	\end{center}
	\caption{Schematic diagrams characterizing the thermodynamic topology of the regular black hole from pure gravity with $\alpha_n = n\alpha^{n-1}$ for the case of $D=5$ and $\alpha=0.1$.
	(a) The $r_+-\tau$ diagram. The left side of the inflection point (marked with black dots) indicates no black hole state, and the right side indicates the presence of two black hole states. The black dashed line represents the lower bound of $r_+$; values less than this will result in $T<0$.
	(b) $\Theta-r_+$ diagram. The red arrows depict the unit vector field $n$ on a section of the $\Theta-r_+$ plane. Here $r_0$ is an arbitrary length scale set by the size of a cavity surrounding the black hole. The zero points ($ZP^a_1$ and $ZP^a_2$) are indicated by black dots.  The green contour $C^a_1$ and the blue contour $C^a_2$ are two closed loops surrounding the zero points. And the black contour $C^a$ representing the closed loop surrounding the entire global structure.
	(c) The deflection angle $\Omega$ as a function of $\vartheta$ for contours $C^a_{1}$ (green), $C^a_{2}$ (blue) and $C^a$ (black). }
	\label{FIG3}
\end{figure}

\begin{figure}[H]
		\begin{center}
		\subfigure[\label{FIG4a}]{\includegraphics[width=5cm]{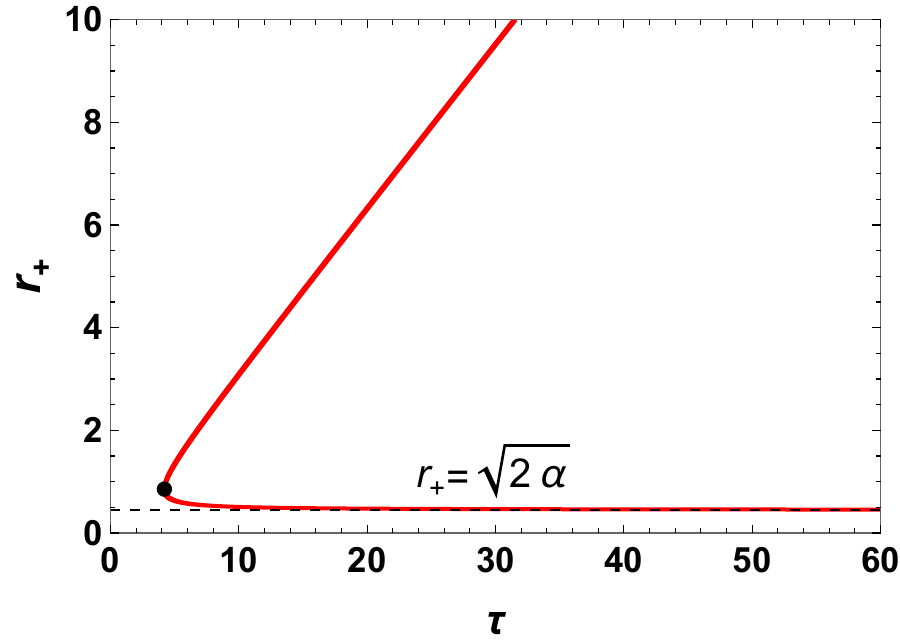}}		
		\subfigure[\label{FIG4b}]{\includegraphics[width=5cm]{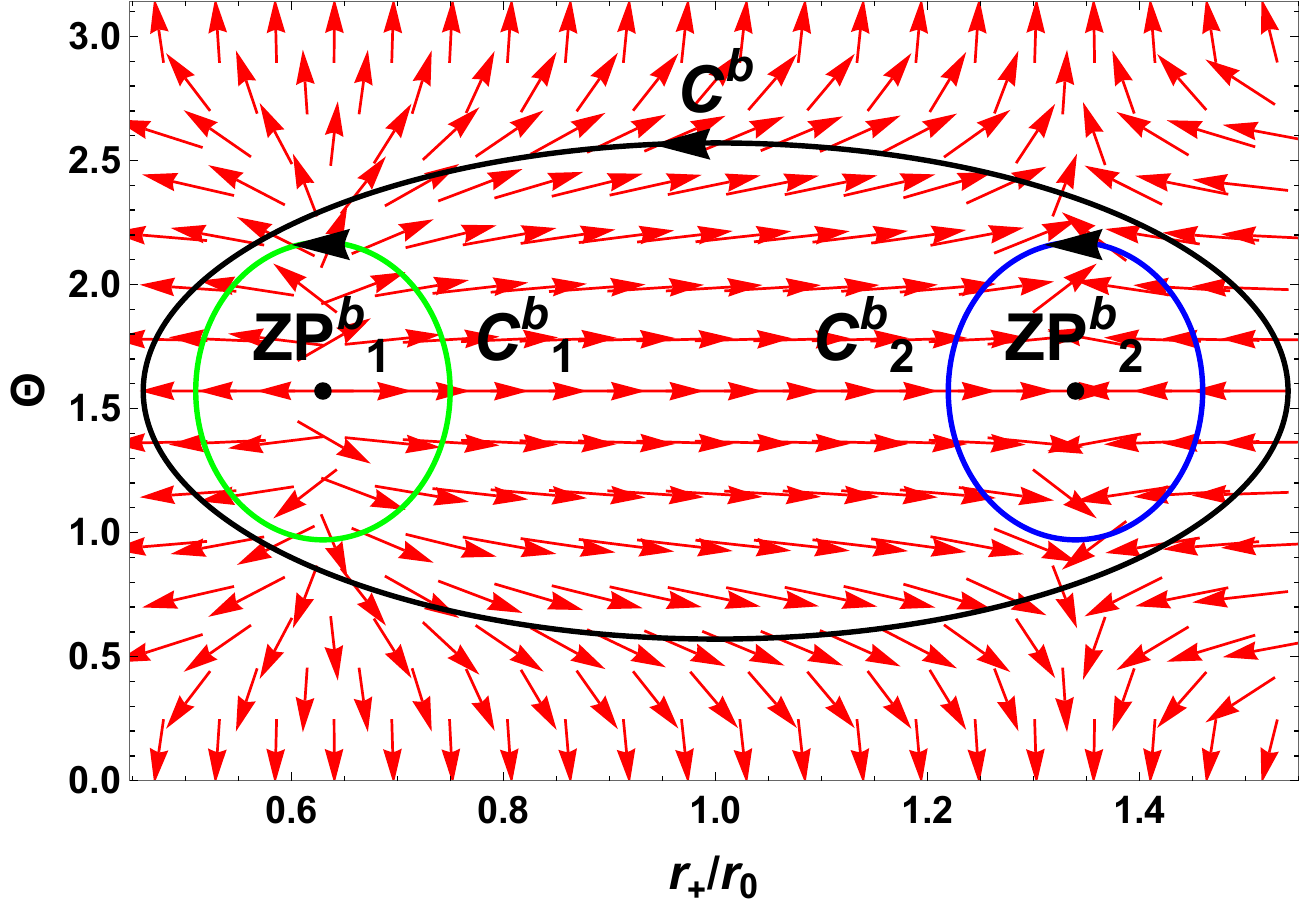}}
		\subfigure[\label{FIG4c}]{\includegraphics[width=5cm]{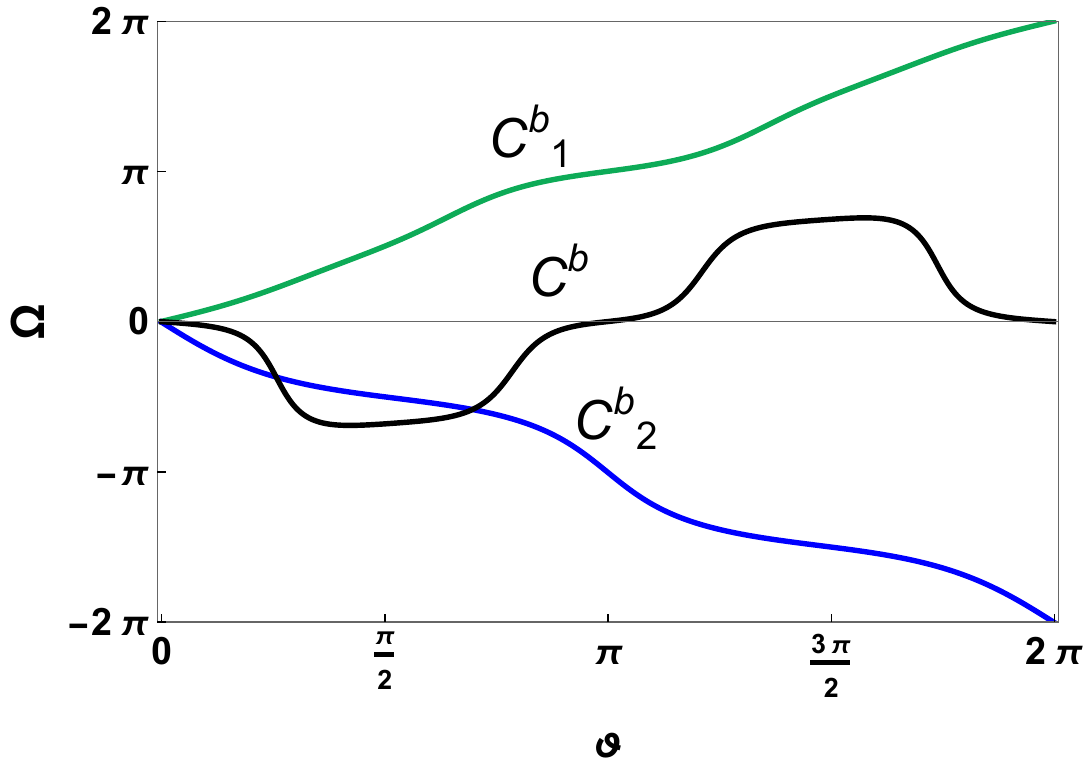}}
	\end{center}
	\caption{Schematic diagrams characterizing the thermodynamic topology of the regular black hole from pure gravity with $\alpha_n = n\alpha^{n-1}$ for the case of $D=7$ and $\alpha=0.1$.
		(a) The $r_+-\tau$ diagram. The left side of the inflection point (marked with black dots) indicates no black hole state, and the right side indicates the presence of two black hole states. The black dashed line represents the lower bound of $r_+$; values less than this will result in $T<0$.
		(b) $\Theta-r_+$ diagram. The red arrows depict the unit vector field $n$ on a section of the $\Theta-r_+$ plane. Here $r_0$ is an arbitrary length scale set by the size of a cavity surrounding the black hole. The zero points ($ZP^b_1$ and $ZP^b_2$) are indicated by black dots.  The green contour $C^b_1$ and the blue contour $C^b_2$ are two closed loops surrounding the zero points. And the black contour $C^b$ representing the closed loop surrounding the entire global structure.
		(c) The deflection angle $\Omega$ as a function of $\vartheta$ for contours $C^b_{1}$ (green), $C^b_{2}$ (blue) and $C^b$ (black).}
	\label{FIG4}
\end{figure}

Taking $\alpha_n=n\alpha^{n-1}$ and substituting it into Eq.~\eqref{eq_h_psi_series}, one will find
	\begin{equation}
		h(\psi)=\frac\psi{(1-\alpha\psi)^2}.
	\end{equation}
Then, the metric function reads
	 \begin{equation}
		f(r)=1-\frac{2mr^{2}}{r^{D-1}+2\alpha m+\sqrt{r^{2(D-1)}+4\alpha mr^{D-1}}}.
	\end{equation}
Through Eqs.~\eqref{eq_M},~\eqref{eq_T}, and~\eqref{eq_S}, we obtain the following thermodynamic quantities
\begin{eqnarray}
		M&=&\frac{(D-2)\pi^{\frac{D-3}{2}}r_+^{D+1}}{8(r_+^2-\alpha)^2\Gamma(\frac{D-1}{2})},
		\label{eq_M1}\\
		T&=&-\frac{3r_+^2-Dr_+^2+\alpha+D\alpha}{4\pi r_+^3+4\pi\alpha r_+},
		\label{eq_T1}\\
		S&=&\frac{\pi^{\frac{D-1}{2}}r_+^D\left(-(D-2)r_+^4+D(r_+^2-\alpha)^2\ _2F_1[2,1-\frac{D}{2};2-\frac{D}{2};\frac{\alpha}{r_+^2}]\right)}{4(r_+^3-\alpha r_+)^2\Gamma(\frac{D-1}{2})}
		\label{eq_S1}.
\end{eqnarray}
It is worth recalling that, using Eq.~\eqref{C}, we determine that the lower bound for $r_+$ is $\sqrt{\alpha}$, which is constrained by the convergence radius of $h(\psi_+)$. Furthermore, from Eq.~\eqref{eq_T1}, it is straightforward to see that this regular black hole becomes extremal when $r_+=\frac{\sqrt{D+1}\sqrt{\alpha}}{\sqrt{D-3}}$. By comparing these two lower bounds, it is clear that for $D\geq5$, this expression provides the true lower bound for $r_+$, avoiding the cases of negative temperature. Thus, the horizon radius $r_+$ lies in the range $\left(\frac{\sqrt{D+1}\sqrt{\alpha}}{\sqrt{D-3}},\infty\right)$.
In Fig.~\ref{FIG3a} and Fig.~\ref{FIG4a}, we present our results in the $r_+-\tau$ graphs, specifically considering $\alpha=0.1$ for $D=5$ and $D=7$, respectively. The lower bounds of $r_+$ are described by black dashed lines. As $r_+$ approaches this bound, $\tau$ diverges to infinity. When $r_+$ increases from its lower bound, $\tau$ decreases to a turning point, corresponding to the condition $\frac{\partial T}{\partial r_+ }=0$, marked by black dots in the figures. Beyond this point, as $r_+$ continues to increase, $\tau$ gradually increases again, eventually diverging to infinity as $r_+\rightarrow\infty$. This behavior is consistent with the properties we have previously analyzed.

Based on Eqs.~\eqref{eq_M1},~\eqref{eq_S1}, and~\eqref{eq_F}, we can derive the expression for the free energy
\begin{equation}
	F=\frac{\pi^{\frac{D-3}{2}} r_+^D \left[(D-2) r_+^3 (2 \pi r_+ + \tau) - 2 D \pi (r_+^2 - \alpha)^2 \, \ _2F_1[2,1-\frac{D}{2};2-\frac{D}{2};\frac{\alpha}{r_+^2}] \right]}{8 (r_+^3 - \alpha r_+)^2 \tau \Gamma\left(\frac{D-1}{2}\right)},
\end{equation}
and the components of the vector $\phi$~\eqref{phi} can be calculated as
\begin{align}
	\phi^{r_+} &= -\frac{\left[ (-2 + D) \pi^{\frac{D-3}{2}} r_{+}^D \left[ 4 \pi r_{+} \left( r_{+}^2 + \alpha \right) - (-3 + D) r_{+}^2 \tau + (1 + D) \alpha \tau \right] \right]}{8 \left( r_{+}^2 - \alpha \right)^3 \tau \Gamma\left( \frac{1}{2} (-1 + D) \right)}\label{phirplus1},\\
	\phi^{\Theta} &= -\cot \Theta \csc \Theta \label{phiTheta1}.
\end{align}

Using the conclusions from Eq.~\eqref{phirplus1} and Eq.~\eqref{phiTheta1}, we can locate and represent the zeros of $\frac{\partial F}{\partial r_+}$, employing the vector $\phi$ field diagram for characterization. We select values of $\tau$ on the right of the turning point in Fig.~\ref{FIG3a} ($r_+-\tau$ graph), ensuring that for a given $\tau$, the $r_+-\tau$ plot
has two intersections, thereby guaranteeing two zeros in the $\Theta-r_+$ plot. The two zeros, labeled $ZP^a_1$ and $ZP^{a}_{2}$ from left to right, are marked by black dots in Fig.~\ref{FIG3b}. Here $r_0$ is an arbitrary length scale set by the size of a cavity surrounding the black hole. To determine the winding number of the zero point, we need to construct a closed loop around it and count the changes in the direction of the vector. For this purpose, we parametrize the closed loop $C^a$ using the angle $\vartheta$, as follows:
\begin{eqnarray}
	\left\{
	\begin{aligned}
		r_+/r_0&=a\cos\vartheta+r_c, \\
		\Theta&=b\sin\vartheta+\frac{\pi}{2}.
	\end{aligned}
	\right.\label{pfs}
\end{eqnarray}
As $\vartheta$ varies from $0$ to $2\pi$, one traces a counterclockwise path around the zero point along this closed loop. To track the change in the direction of the vector, it is useful to define the deflection angle~\cite{Wei:2020rbh}
\begin{equation}
\Omega(\vartheta)=\oint_C \epsilon_{ab}n^a\partial_{i}n^bdx^{i}.
\end{equation}
Then the winding number shall be

\begin{equation}
W=\Omega(2\pi)/2\pi.
\end{equation}

Using the method described above, by examining the deflection angle along the green circle $C^a_1$ and the blue circle $C^a_2$ shown in Fig.~\ref{FIG3c}, we determine that the winding number for $ZP^a_1$ is $W_1=1$, indicating a thermodynamical stable black hole state, while for $ZP^a_2$, it is $W_2=-1$, indicating an unstable black hole state. These support that such topological class is $W^{0+}$ rather than $W^{0-}$. The black circle $C^a$ represented by the black line surrounds the entire global structure shown in Fig.~\ref{FIG3c}, leads to the winding number $W=0$. It is noteworthy that if we choose a $\tau$ to the left of the turning point, where no intersection points exist in the $r_+-\tau$ graph, there will be no zeros in the $\Theta-r_+$ graph. Certainly, by summing the topological numbers of all the local zeros, we can also obtain the total topological number, which is $W = -1 + 1 = 0$. Using the same method, we can obtain the results for $D=7$ and $\alpha=0.1$, with similar results shown in Figs.~\ref{FIG4a}, \ref{FIG4b}, and \ref{FIG4c}.

Certainly, we can discuss the cases corresponding to different values of $\alpha$ in various dimensions, all of which show consistent results, though we will not present all of them here. It is quite natural to observe this, as our previous discussions were based on general conditions, and the values of $D$ and $\alpha$ do not affect our conclusions. In other words, when $\alpha$ and $D$ vary independently, they do not affect the asymptotic behavior of $\tau$ or $\frac{\partial F}{\partial r_+}$.  Regardless of how $\alpha$ and $D$ vary, this regular black hole consistently exhibits a topological number of $W=0$, indicating that black holes in different dimensions exhibit the same topological behavior. Therefore, this result supports our earlier conclusions.


\subsection{Dymnikova black hole from corrections}

The Dymnikova black hole can be interpreted as a solution to the Einstein equations (in four or higher dimensions) with an energy-momentum tensor describing matter that exhibits exponential decay \cite{Dymnikova:1992ux,Paul:2023pqn}. Konoplya and Zhidenko derived this solution using the pure gravity approach in Ref.~\cite{Konoplya:2024kih}. Although the expression of $h(\psi)$ differs from that in the previous subsection, detailed analysis yields similar results.
		
The function $h(\psi)$ for this Dymnikova black hole reads~\cite{Konoplya:2024kih}
\begin{equation}
		h(\psi)=\frac{\psi}{1+\alpha\psi \L\left(-\frac{e^{-1/\alpha\psi}}{\alpha\psi}\right)},\quad\alpha>0,
		\label{h(psi)4}
\end{equation}
where $\L(x)$ is the principal Lambert function, which fulfills the equation
\begin{equation}
			\L(x)e^{\L(x)}=x.
			\label{Lambert}
\end{equation}
Then the metric function is
\begin{equation}
			f(r)=1-\frac{\mu}{r^{D-3}}\left(1-e^{-\cfrac{r^{D-1}}{\alpha\mu}}\right).
\end{equation}
We can naturally obtain the corresponding thermodynamic quantities by Eqs.~\eqref{eq_M},~\eqref{eq_T}, and~\eqref{eq_S},
\begin{eqnarray}
			M&=&\frac{(D-2)\pi^{\frac{D-3}{2}}r_+^{D-1}}{8\Gamma(\frac{D-1}{2})\left(r_+^2+\alpha A\right)},	 \label{eq_M2}\\
			 T&=&\frac{\left(D-3\right)+(D-1)A}{4\pi r_+},
			\label{eq_T2}\\ S&=&\frac{\left(D-2\right)\pi^{\frac{D-1}{2}}}{4\Gamma(\frac{D-1}{2})}\int\frac{2r_+^{D-1}}{\left(1+A\right)\left(r^{2}+\alpha A\right)}\mathrm{d}r_+	\label{eq_S2},
\end{eqnarray}
where
\begin{equation}
	A=\L\left(-\frac{exp\left(-{\frac{r_+^2}{\alpha}}\right)r_+^2}{\alpha}\right). \nonumber
\end{equation}

\begin{figure}[H]
		\begin{center}
			\subfigure[\label{FIG5a}]{\includegraphics[width=5cm]{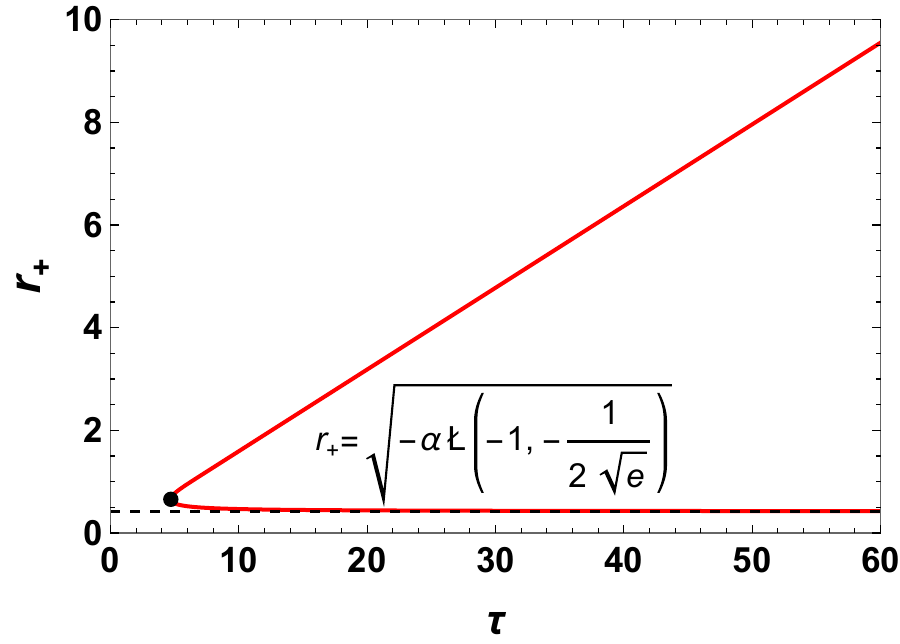}}
			\subfigure[\label{FIG5b}]{\includegraphics[width=5cm]{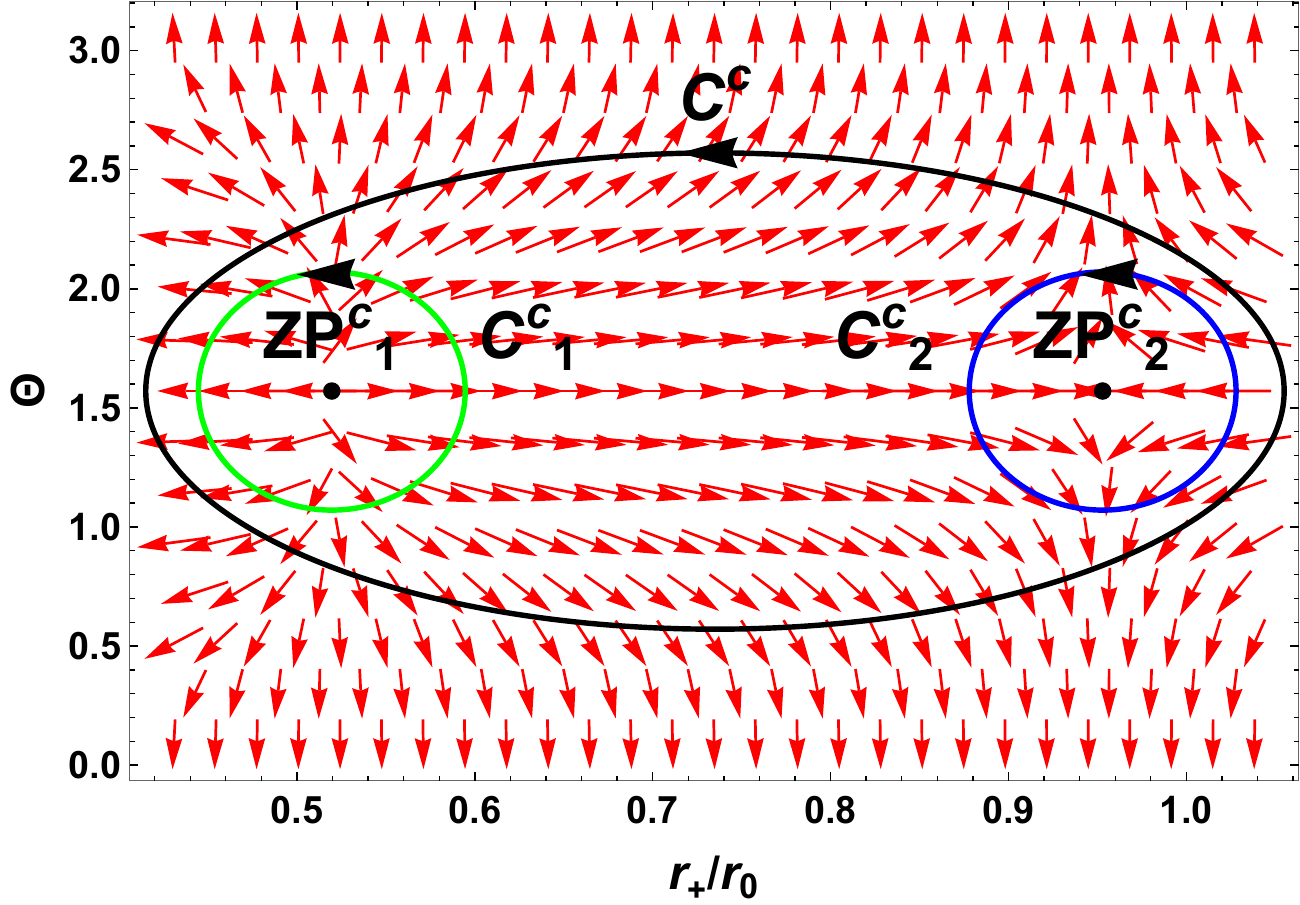}}
			\subfigure[\label{FIG5c}]{\includegraphics[width=5cm]{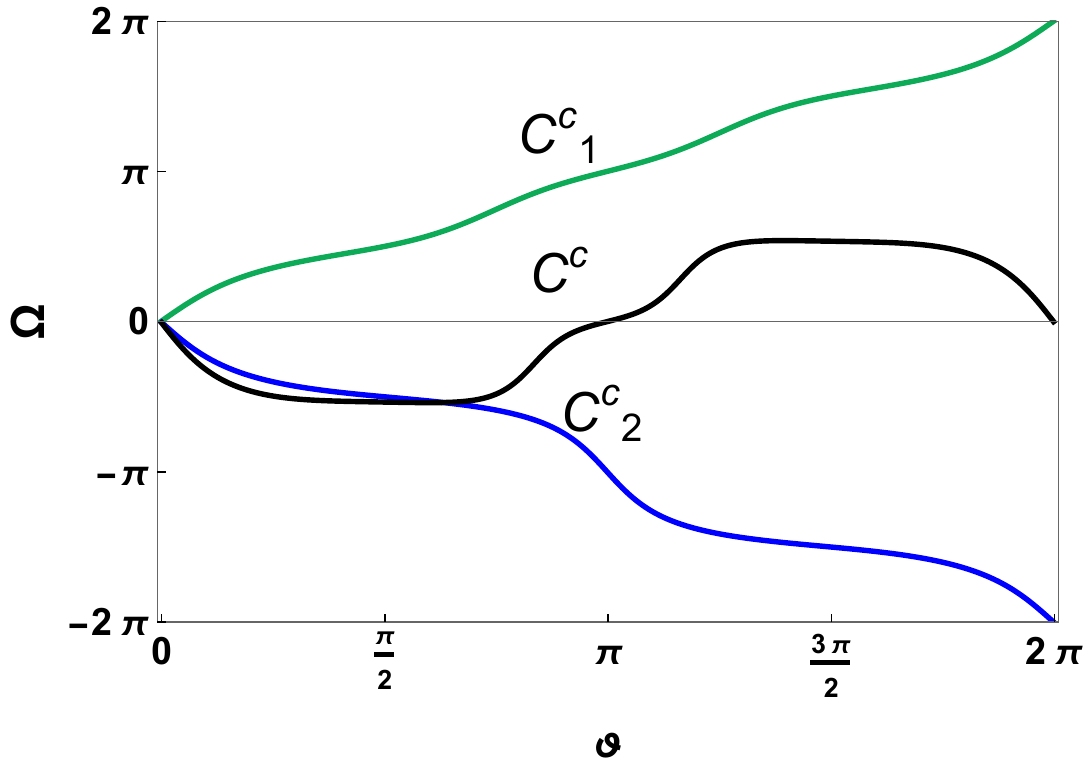}}
	\end{center}
\caption{Schematic diagrams characterizing the thermodynamic topology of the of the Dymnikova black hole for the case of $D=5$ and $\alpha=0.1$.
	(a) The $r_+-\tau$ diagram. The left side of the inflection point(marked with black dots) indicates no black hole state, and the right side indicates the presence of two black hole states. The black dashed line represents the lower bound of $r_+$; values less than this will result in $T<0$.
	(b) $\Theta-r_+$ diagram. The red arrows depict the unit vector field $n$ on a section of the $\Theta-r_+$ plane. Here $r_0$ is an arbitrary length scale set by the size of a cavity surrounding the black hole. The zero points ($ZP^c_1$ and $ZP^c_2$) are indicated by black dots.  The green contour $C^c_1$ and the blue contour $C^c_2$ are two closed loops surrounding the zero points. And the black contour $C^c$ representing the closed loop surrounding the entire global structure.
	(c) The deflection angle $\Omega$ as a function of $\vartheta$ for contours $C_{1}^c$ (green), $C_{2}^c$ (blue) and $C^c$ (black).  }
\label{FIG5}
	\end{figure}

\begin{figure}[H]
		\begin{center}
			\subfigure[\label{FIG6a}]{\includegraphics[width=5cm]{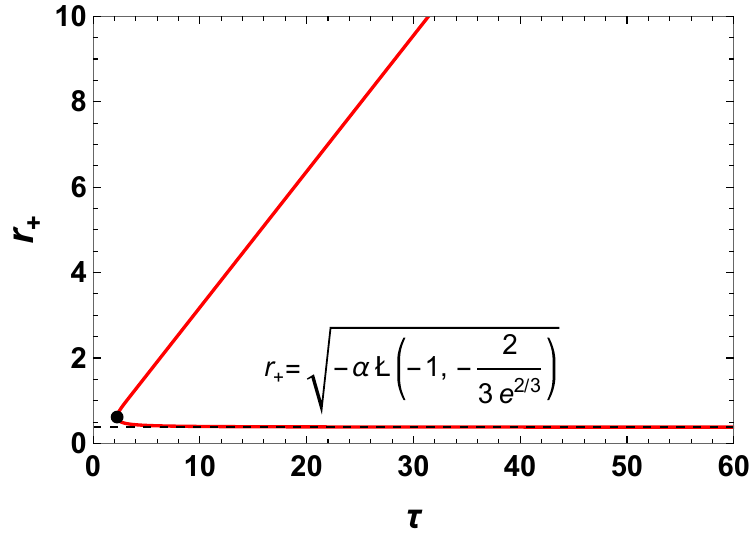}}
			\subfigure[\label{FIG6b}]{\includegraphics[width=5cm]{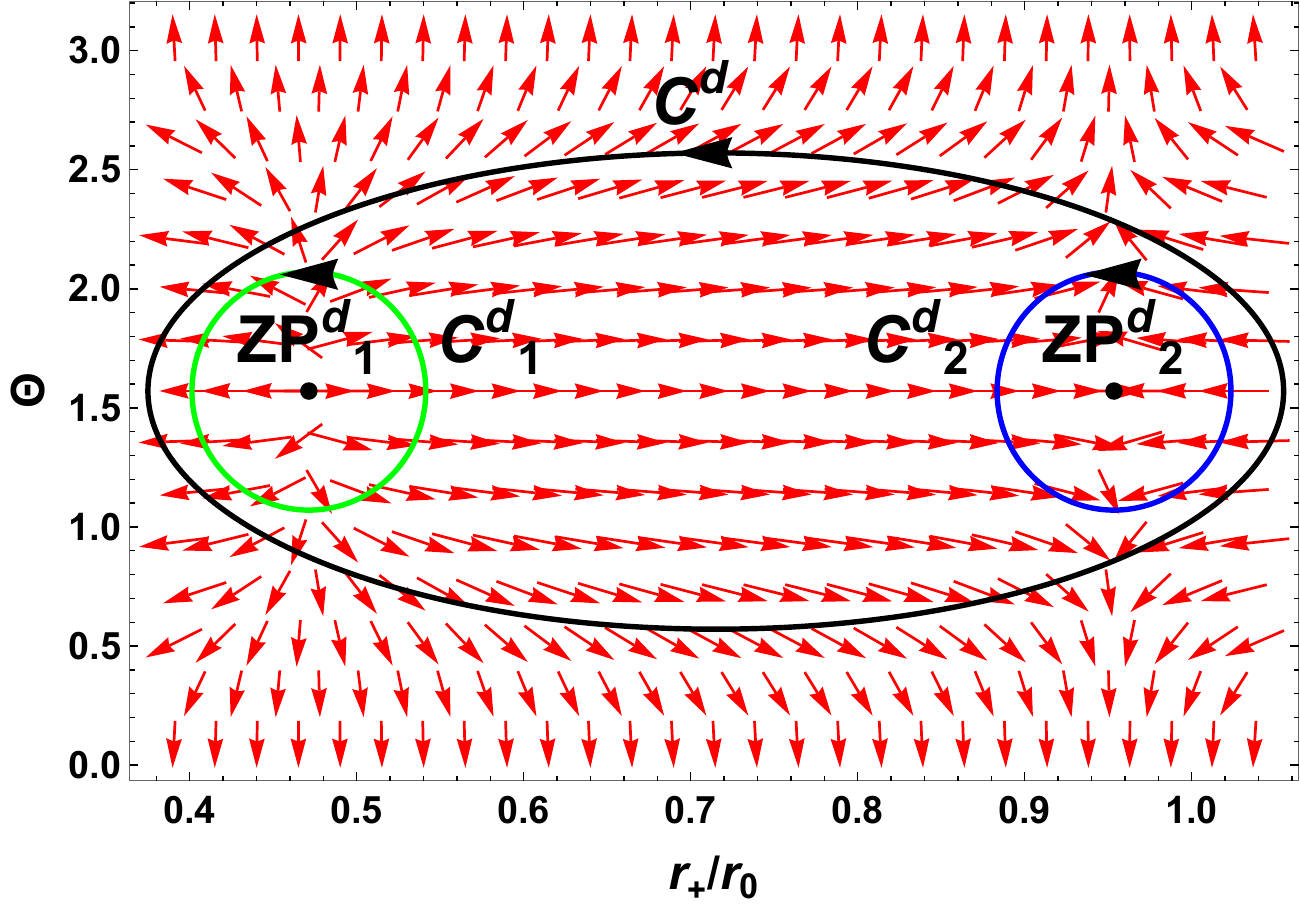}}
			\subfigure[\label{FIG6c}]{\includegraphics[width=5cm]{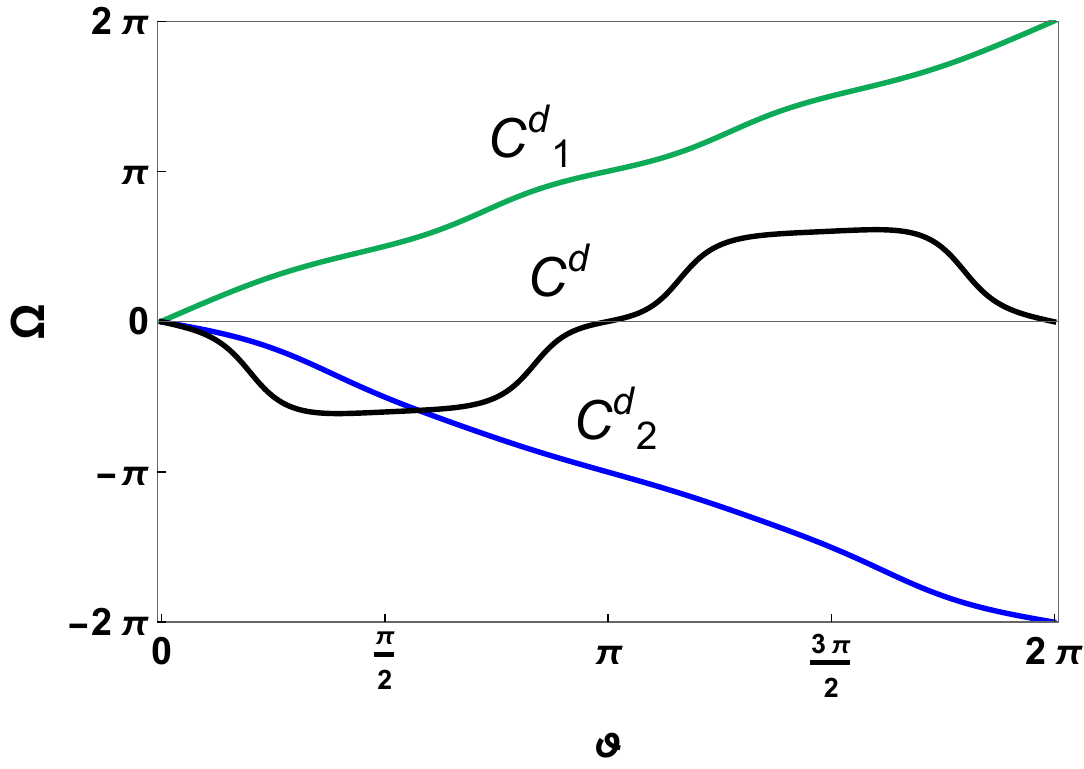}}
		\end{center}
		\caption{Schematic diagrams characterizing the thermodynamic topology of the of the Dymnikova black hole for the case of $D=7$ and $\alpha=0.1$.
			(a) The $r_+-\tau$ diagram. The left side of the inflection point(marked with black dots) indicates no black hole state, and the right side indicates the presence of two black hole states. The black dashed line represents the lower bound of $r_+$; values less than this will result in $T<0$.
			(b) $\Theta-r_+$ diagram. The red arrows depict the unit vector field $n$ on a section of the $\Theta-r_+$ plane. Here $r_0$ is an arbitrary length scale set by the size of a cavity surrounding the black hole. The zero points ($ZP^d_1$ and $ZP^d_2$) are indicated by black dots.  The green contour $C^d_1$ and the blue contour $C^d_2$ are two closed loops surrounding the zero points. And the black contour $C^d$ representing the closed loop surrounding the entire global structure.
			(c) The deflection angle $\Omega$ as a function of $\vartheta$ for contours $C^d_{1}$ (green), $C^d_{2}$ (blue) and $C^d$ (black). }
\label{FIG6}
	\end{figure}

Similar to the previous section, it is straightforward to verify that the lower bound for $r_+$, determined by the convergence radius of the function $h(\psi)$, is $\sqrt{\alpha}$. To avoid negative temperatures, we obtain $r_{+\text{min}}$, and the event horizon must lie above this value. While providing a general expression is challenging, we illustrate the $r_{+\text{min}}$ for different cases using dashed black lines in Fig.~\ref{FIG5a} and Fig.~\ref{FIG6a}. By substituting specific values, it is easy to verify that $r_{+\text{min}}$ is always greater than $\sqrt{\alpha}$. Thus, the lower bound derived from the zero-temperature condition is indeed the true lower bound for $r_+$.

Based on Eqs.~\eqref{eq_M2},~\eqref{eq_S2}, and~\eqref{eq_F}, we can derive the expression for the free energy
\begin{equation}
	F=\frac{(-2+D)\pi ^{\frac{-3+D}{2}}}{8\Gamma \left( \frac{-1+D}{2} \right)}
	\left[\ \frac{-4\pi }{\tau}\int{\frac{r_{+}^{D-1}}{\left( 1+A \right) \left( r_{+}^{2}+\alpha A \right)}\,\mathrm{d}r_+}+\frac{r_{+}^{D-1}}{r_{+}^{2}+\alpha A} \right] ,
\end{equation}
and the components of the vector $\phi$ can be calculated as
\begin{align}
		\phi^{r_+} &= \frac{(-2 + D) \pi^{\frac{-3 + D}{2} } r_+^{-2 + D} \left[-4 \pi r_+ + (-3 + D) \tau + (-1 + D) \tau \, A\right]}{8 \tau \Gamma\left(\frac{-1 + D}{2} \right) \left(1 + A\right) \left(r_+^2 + \alpha A\right)} , \\
	\phi^{\Theta} &= -\cot \Theta \csc \Theta.
\end{align}
Using the same approach as in the previous subsection, we can obtain the corresponding $r_+-\tau$ diagram, the vector field $\phi$, and the deflection angle shown in Fig.~\ref{FIG5} and Fig.~\ref{FIG6}. The cases for $\alpha = 0.1$, $D = 5$, and $D = 7$ are considered separately. Similarly, we can examine cases for different values of $\alpha$ across various dimensions, all of which yield consistent results, although we will not present every case here. The variations in the two parameters, $\alpha$ and $D$, do not affect the asymptotic behavior of $\tau$ or $\frac{\partial F}{\partial r_+}$, ultimately leading to a topological number $W=0$, which further confirms our analysis.

\section{Discussion and Conclusion}\label{Sec_Disscusion}

In this paper, we analyzed the thermodynamical topology of regular black holes arising from pure gravity from a general perspective. We found that regular black holes constructed using the method from Ref.~\cite{Bueno:2024dgm} exhibit certain similar thermodynamic properties. We also observed that these black holes, with at most two horizons, have similar topological properties and the same topological number, $W=0$, as revealed by the asymptotic behavior of the constructed vector. In particular, they belong to $W^{0+}$ topological classification given in Ref. \cite{Wei24}, through which one can find that the the innermost and outermost black hole states are local thermodynamical stable and unstable. Given the general nature of our discussion, this result suggests that the thermodynamic topological property is intrinsically linked to the construction method of these regular black holes.

Specifically, we consider two representative examples of regular black hole solutions: one characterized by $\alpha_n = n\alpha^{n-1}$ and the Dymnikova black holes. We demonstrated the zero points of the free energy by constructing a vector field $\phi=\begin{pmatrix}\frac{\partial F}{\partial r_+},-\cot\Theta\csc\Theta\end{pmatrix}$. Through detailed calculations, we observed that for small $\tau$, there are no black hole states, leading to a vanishing topological number. However, for large $\tau$, two zero points of the constructed vector field emerge, corresponding to the regular black hole states. We then enclosed these zeros with a closed loop and counted the changes in the vector's direction to determine the winding number associated with each zero point. Consequently, we obtained the overall global winding number, which yields a topological number of $W = 0$. This result further supports our conclusion that the regular black holes (with at most two horizons in pure gravity) belong to the same topological class $W^{0+}$. Moreover, the small and large black hole states are stable and unstable indicated from their winding numbers, which strongly supports the thermodynamical properties of $W^{0+}$. Although we only provide a few specific examples, this does not affect the generality of our conclusion, as the choice of specific parameters does not alter the results of our previous discussion on the asymptotic behavior of $\tau$ and $\frac{\partial F}{\partial r_+}$.

On the other hand, our discussion has certain limitations. We only considered the case where the Hawking temperature has a single zero point and did not address more complex scenarios involving multiple horizons. If $\alpha_{n}$ is chosen as an appropriate function, such cases may arise, requiring more detailed study. Nevertheless, for the regular black hole solutions constructed by this method that have been identified so far, our findings remain applicable. Furthermore, the reasons behind the topological consistency of these black holes, and whether there are deeper underlying factors in their construction methods, merit further reflection and discussion.

\appendix
\section{Behavior of Hawking temperature at the lower bound $r_+ = \sqrt{C}$}
		\label{appendixprove}

We provide a simple proof for the conclusion in Sec.~\ref{Sec_GeneralDisscusion} about Eq.~\eqref{eq_T} or~\eqref{eq_T_subsitutingh(psi)2} in the main text, namely, that when $r_+$ is sufficiently close to $\sqrt{C}$, the temperature $T$ can become negative. Before proceeding, we first present a claim: for smooth function $f: (a,b) \rightarrow \mathbb{R}$ satisfying the condition,
	\begin{equation}
		\lim_{x \rightarrow b^{-}} f(x) = + \infty,
	\end{equation}
	it is always possible to find a sequence $ \Xi = \left\{ \xi _n, n\in \mathbb{N} ^+ \middle| a<\xi _n<b, \xi _n\le \xi _{n+1}, \forall n\in \mathbb{N} ^+ \right\} $ satisfying $ \lim_{n\rightarrow \infty} \xi_n = b$ such that $\lim_{n \rightarrow \infty} 1/f'(\xi_n) =0$.

We presents a simple proof for this claim. Since the left-sided limit of $f(x)$ at $x = b$ is infinity, $f(x)$ can exceed any number if $x$ is sufficiently close to $b$. Therefore, we can always find such a sequence  $\left\{ x_n,n\in \mathbb{N} ^+ \middle| a<x_n<x_{n+1}<b, f\left( x_n \right) <f\left( x_{n+1} \right) , \forall n\in \mathbb{N} ^+; \lim_{n\rightarrow \infty} x_n = b \right\}
	$. Now we construct the sequence $\Xi$ by finding the value of $x$ in the closed interval $[x_1, x_n]$ that maximizes $f'(x)$, and let one of these values be $\xi_{n-1}$. This implies the following inequality holds,
\begin{equation}
		f(x_n) \leq f(x_1) + f'(\xi_{n-1}) (x_n - x_1) < f(x_1) + f'(\xi_{n-1}) (b - x_1), \ \forall n \ge 2.
\end{equation}
	It is evident that the constructed $\xi_n$ satisfies $a < \xi_n \leq \xi_{n+1} < b, \forall n \in \mathbb{N}^+$. This indicates that the sequence $\Xi$ has a limit, denoted as $\xi_\infty$. Furthermore, it can be shown that $f'(\xi_n) \leq f'(\xi_{n+1}), \forall n \in \mathbb{N}^+$. Therefore, if $f'(\xi_\infty)$ is finite, we have
	\begin{equation}
		f(x_n) < f(x_1) + f'(\xi_{n-1}) (b - x_1) \leq f(x_1) + f'(\xi_\infty) (b - x_1), \ \forall n \ge 2.
	\end{equation}
	This contradicts $\lim_{x \rightarrow \infty} f(x_n) = \infty$. Hence, $\lim_{x \rightarrow \infty} f'(x_n) = f'(\xi_\infty) = \infty$. Since $f'(x)$ is finite for any $x \in (a, b)$, we have $\xi_\infty = b$. Thus, the proof of this claim is complete.

	In Sec.~\ref{Sec_GeneralDisscusion}, $h(\psi_+)$ is a smooth function on the interval $(0, 1/C)$, and $\lim_{x \rightarrow (1/C)^{-}} h(\psi_+) =+ \infty$. This implies that $\ln(h(\psi_+))$ also exhibits the same property. Our claim asserts that there exists a sequence  $ \Xi = \left\{ \xi _n, n\in \mathbb{N} ^+ \middle| 0<\xi _n<1/C, \xi _n\le \xi _{n+1}, \forall n\in \mathbb{N} ^+ \right\} $ satisfying $\lim_{n \rightarrow \infty} \xi_n = 1/C$, such that
	\begin{equation}
		\lim_{n \rightarrow \infty} \frac{h(\xi_n)}{h'(\xi_n)} = 0.
		\label{eq_hxi}
	\end{equation}
	For the temperature,
	\begin{equation}
		\lim_{n \rightarrow \infty} T(\xi_n) = - \frac{1}{2 \pi \sqrt{C}}.
	\end{equation}
	It is noteworthy that Eq.~\eqref{eq_hxi} implies that if $\lim_{\psi_+ \rightarrow (1/C)^-} \frac{h(\psi_+)}{h'(\psi_+)}$ exists, then its result must be zero, which strengthens all the conclusions. Nevertheless, we can ascertain that near the lower bound $r_+ = \sqrt{C}$, there exists an $r_+$ for which the temperature is negative. This, combined with the discussion in Sec.~\ref{Sec_GeneralDisscusion}, indicates that the temperature must have at least one zero point.

\section*{Acknowledgments}
This work was supported by the National Natural Science Foundation of China (Grants No. 12475055, No. 12075103, and No. 12247101).

\end{document}